\documentclass[reprint,
superscriptaddress,
frontmatterverbose, 
amsmath,amssymb,
aps,
prb,
floatfix,
showkeys
]{revtex4-2}

\usepackage{float}
\usepackage{graphicx}
\usepackage{dcolumn}
\usepackage{bm}
\usepackage[left=2cm,right=2cm,top=2cm,bottom=2cm]{geometry}
\usepackage{mathtools}
\usepackage{amsmath}
\usepackage{lipsum}
\usepackage{xcolor}
\usepackage{soul}
\usepackage{multirow}
\usepackage{ulem}
\mathchardef\hyphen="2D

\usepackage{booktabs}
\usepackage{siunitx}
\graphicspath{{./}}

\begin{document}

\title{Exact heat flux formula and its spectral decomposition in molecular dynamics for arbitrary many-body potentials}

\author{Markos Poulos}
\email[Corresponding Author: ]{markos.poulos@insa-lyon.fr}
\affiliation{INSA Lyon, CNRS, CETHIL, UMR5008, 69621 Villeurbanne, France}

\author{Donatas Surblys}
\affiliation{Institute of Fluid Science, Tohoku University, 2-1-1 Katahira, Aoba-ku, Sendai 980-8577, Japan }
 
\author{Konstantinos Termentzidis}
\affiliation{CNRS, INSA Lyon, CETHIL, UMR5008, 69621 Villeurbanne, France}

\begin{abstract}
In this study we have derived an exact framework for the calculation of the heat flux and its spectral decomposition in Molecular Dynamics (MD) for arbitrary many-body potentials. This work addresses several lacks and limitations of previous approaches and allows for the accurate computational study of thermal properties in a wide variety of many-body systems with MD. We have tested our modifications with Green-Kubo (GK) and Non-Equilibrium MD (NEMD) simulations for various 2D and 3D material systems using the Tersoff and Stillinger-Weber potentials as examples. The spectral decomposition of the heat current was also calculated for monolayer graphene (1LG) and MoS$_2$, for different system lengths. Our results show that the heat current calculated by our method is consistently in agreement with the thermostat current in NEMD, while previous implementations can estimate quite poorly the thermal conductivity both under GK and NEMD simulations, and both for 2D and 3D materials. The decomposition of the heat current also sheds light on the contribution of different phonon modes to thermal conductivity and its dependence on length. Our methodology is implemented in the widely used LAMMPS code specifically for the Tersoff and SW potentials, and it is readily applicable to the vast majority of many-body MD potentials.

\end{abstract}

\maketitle

\section{\label{sec:Intro}Introduction:}
Molecular Dynamics (MD) simulations are a powerful tool for studying the thermal properties of materials in the nanoscale~\cite{Oliveira2019,Kinaci2012,Wei2019,Termentzidis2018}. Within MD there are two principal and widely used methods for calculating thermal properties, namely the Green-Kubo (GK) method and the Non-Equilibrium MD (NEMD) method. The GK method is a statistical technique that uses the autocorrelation function of the instantaneous heat flux of a system in equilibrium to calculate the thermal conductivity $\kappa$ from the fluctuation-dissipation theorem~\cite{Green1954,Kubo1957}, while NEMD is based on the application of a temperature gradient to a system by means of connecting thermal reservoirs at a temperature difference $\Delta T$. The resulting heat flux $\boldsymbol{J}$ is calculated and, using Fourier's law, thermal conductivity is determined as $\kappa = -\boldsymbol{J}/\boldsymbol{\nabla} T$ (or sometimes seen with the gradient replaced by the average temperature slope $\Delta T/L$ where $L$ is the system length~\cite{Hu2020}).

In both methods, the heat flux (HF) is a central quantity needed in order to calculate $\kappa$. For NEMD simulations, the HF is usually calculated by measuring the energy exchange rate between the thermostat regions and the rest of the system, while for GK the time-dependent HF is calculated as an average over the system volume from the instantaneous atomic velocities and forces acting on the atoms (see Section~\ref{sec:Methods} for details). The instantaneous heat current $Q^{1 \rightarrow 2}$ flowing across an interface separating the system into regions $1$ and $2$ is also an important quantity, as it is used to calculate the Interfacial Thermal Conductance (ITC) of interfaces and heterostructures~\cite{Suarez2018, Hu2011, Merabia2014}. $Q^{1 \rightarrow 2}$ can also be decomposed to reveal the contribution of phonon modes of specific energy both to the the thermal conductivity and ITC, known as the Spectral Decomposition of the Heat Current (SDHC)~\cite{Saaskilahti2014, Saaskilahti2015, Saaskilahti2016,Chalopin2013}. The calculation of $Q^{1 \rightarrow 2}$ and its spectral decomposition requires a definition of the mechanical power exerted by the interatomic forces that is due only to atoms from across the interface, which is trivial for pair potentials~\cite{Saaskilahti2014}, but not evident for many-body interactions.

LAMMPS is a widely used open-source MD code, employing a vast variety of interatomic potentials to simulate various kinds of interactions (covalent, ionic, Van der Waals, etc.), and capable of performing both NEMD and GK simulations ~\cite{Plimpton1995}. For NEMD, the calculation of the HF from the thermostat energy exchange rate is straightforward and a priori exact. However, for equilibrium methods like GK it is based on the calculation of the atomic stress tensor, for which LAMMPS originally used a formula that is valid only for pair potentials. This can lead to incorrect results for $\kappa$. The correct formulation has been known for a long time, since Torii \textit{et al.} theoretically derived exact formulas for calculating the HF both for a control volume and a control surface, for arbitrary many-body potentials and an arbitrary partitioning of the many-body potential energy among the participating atoms~\cite{Torii2008}. 

The correct calculation of $\kappa$ has been demonstrated for some cases, such as molecular systems using simple bond, angle and torsion terms~\cite{Surblys2019, Boone2019}, carbon systems using the Tersoff potential~\cite{fan2015force}, and recently for some Machine Learning potentials~\cite{Hamakawa2025}. In these works, the authors have proposed and published modifications to the LAMMPS HF formula; Surblys~\textit{et al}.~\cite{Surblys2019} and Boone \textit{et al}.~\cite{Boone2019} have published corrections to the atomic stress tensor for bond, angle and torsion terms, while Surblys~\textit{et al} also extended this formalism to systems with constraint dynamics~\cite{Surblys2021}. Hamakawa \textit{et al}. corrected the HF formula for some Machine Learning potentials~\cite{Hamakawa2025}, while Fan \textit{et al}.~\cite{fan2015force} have published an implementation of an exact formula for the calculation of the HF ith the Tersoff, SW and EAM potentials, as well as some Machine Learning potentials, in a code running on GPU's~\cite{fan2015force, Fan2017GPUMD,Fan2025GPUMD}.

Therefore, implementations of exact formulas for the HF in a control volume are available but with limited applicability to specific systems. 
Calculation of $Q^{1 \rightarrow 2}$ through a control surface is also available, but only for pair-wise potentials, while it completely lacks many-body implementations for both $Q^{1 \rightarrow 2}$ and the SDHC. As a result, the SDHC has been calculated in the past only for systems with simple pairwise interactions~\cite{Saaskilahti2016}, or by approximating the many-body forces as pair harmonic forces, under the so-called harmonic approximation~\cite{Chalopin2013, Saaskilahti2014}. Thus in this work we have extended and implemented in LAMMPS the Torii \textit{et al}. formulas~\cite{Torii2008} for calculating the heat flux for both a control volume and a control surface, and subsequently its spectral decomposition, for arbitrary many-body potentials.

The article is organized as follows: In Section~\ref{sec:Methods} we present the theoretical background based on the derivations by Torii \textit{et al}.~\cite{Torii2008} for the calculation of the heat flux in a control volume, the heat current through a control surface, and the Spectral Decomposition of the Heat Current (SDHC). In Section~\ref{sec:Results} we present some results of comparative tests of our implementation with the existing LAMMPS implementation, for GK and NEMD simulations and for various test systems (both 2D and 3D solids) using the Tersoff and the Stillinger-Weber potentials, as well as some results of SDHC calculations for 1L graphene and MoS$_2$. Finally, in Section~\ref{sec:Conclusions} we summarize our findings and highlight the contributions of our work. Computational details about the simulation tests are provided in the Supplementary Material.

\section{\label{sec:Methods} Methods}

\noindent\textbf{Hardy Formula:} The general formula for the instantaneous heat flux $\boldsymbol{J}$ passing through a cross-section of the system has been derived by Hardy in the early 1960s, initially within the framework of quantum mechanics, and subsequently for the classical limit as~\cite{Hardy1963}
\begin{equation}
    \boldsymbol{J} = \frac{1}{V}  \sum_i \boldsymbol{v}_i \epsilon_i + \sum_{i \ne j} \boldsymbol{r}_{ij} \left( \boldsymbol{F}_{ij} \cdot \boldsymbol{v}_j \right)
    \label{eq:HeatFlux_Hardy}
\end{equation}
\noindent where $V$ is the system volume, $\boldsymbol{r}_{ij} = \boldsymbol{r}_i - \boldsymbol{r}_j$ is the interparticle distance, $\boldsymbol{v}_i$ is the velocity of atom $i$, $\epsilon_i$ is the per-atom energy, and the forces $\boldsymbol{F}_{ij}$ are defined in the classical manner as $\boldsymbol{F}_{ij} = -\frac{\partial U_j}{\partial \boldsymbol{r}_i}$, where it is implied that the total many-body potential energy $U$ is partitioned among the atoms as $U=\sum_j U_j$. The choice of the partitioning $U_j$ into atomic sites is arbitrary and non-unique, but it has been shown that the thermal conductivity calculations do not require a specific partition scheme~\cite{Marcolongo2016, Ercole2016, Marcolongo2020}.

In practical implementations of MD simulations, the forces acting on the atoms are calculated from empirical interatomic potentials with terms involving the coordinates of, say, $N$-atoms. A loop is then performed over all different $K_N$ interaction groups containing $N$ atoms, each labeled by the superscript $k$, and for each $k$ term the forces $\boldsymbol{F}^{k}_i$ due to the potential energy term $U^k$ acting on each atom $i$ of the group are calculated. A choice of partitioning of $U^k$ to each participating atom $i$ is also implicitly made in the MD algorithms, usually dictated by the form of the potential function, and which can be represented by a set of weights $p_i^k$, where $p_i^kU^K$ is the fraction of the potential energy $U^k$ of the $N$-body term $k$ that is assigned to atom $i$ in the group $k$. The weights must also satisfy $\sum_i p_i^k=1$.

\subsection{Heat Flux in a Control Volume}
\noindent\textbf{Torii Formula:} Using a similar choice of notation, Torii \textit{et al}. showed that the instantaneous heat flux $J_x$ along direction $x$ averaged over the control volume $V$ of the system is given by~\cite{Torii2008}
\begin{align}
\boldsymbol{J}V &= \sum_{i} \boldsymbol{v}_i E_i+ \sum_{k}^{K_N}\nonumber \\
& \sum_{i,j \in k \atop i>j}\ \left(p_j^k \boldsymbol{F}_i^k \cdot \boldsymbol{v}_i^k - p_i^k \boldsymbol{F}_j^k \cdot \boldsymbol{v}_j^k \right) \left(\boldsymbol{r}_i -\boldsymbol{r}_j\right) \label{eq:HeatFlux_Torii_V}
\end{align}
\noindent where the per-atom energy $E_i$ of atom $i$ partitioned according to the weights $p_i^k$ is given by:
\begin{equation}
E_{i} = \frac{1}{2} m_i v_i^2 +  \sum_{k}^{K_N} \ p_i^k U^k \label{eq:Energy_Torii}
\end{equation}

\noindent The first term in eq.~(\ref{eq:HeatFlux_Torii_V}) is the kinetic (convective) contribution, while the second term is the virial (purely conductive) contribution. The convective part is proportionally small in systems where there is no net flow of particles, and so it is usually neglected in solid systems like the ones that we will examine in this work. In the virial term, the first summation runs over all the different interaction groups $k$, each containing $N$ terms, and the second summation runs over all pairs of atoms $i$ and $j$ in each group $k$.\\

\noindent\textbf{Atomic Stress Tensor (Centroid): }By rearranging terms in order to express eq.~(\ref{eq:HeatFlux_Torii_V}) in a per-atom basis, one can write it more compactly as
\begin{align}
    \boldsymbol{J} &= \boldsymbol{J}^{\text{conv}} + \boldsymbol{J}^{\text{virial}} \\
    \boldsymbol{J}^{\text{conv}}V &=\sum_i E_i \boldsymbol{v}_i \\
    \boldsymbol{J}^{\text{virial}}V &= - \sum_i \overleftrightarrow{\boldsymbol{{\sigma}}}_i \cdot \boldsymbol{v}_i \label{eq:HeatFlux_virial}
\end{align}

\noindent where $\boldsymbol{J}^{\text{conv}}$ and $\boldsymbol{J}^{\text{virial}}$ correspond to the convective and the virial part of the heat flux respectively, and $\overleftrightarrow{\boldsymbol {{\sigma}}_i}$ is the atomic virial stress tensor, given by
\begin{align}
    \overleftrightarrow{\boldsymbol {{\sigma}}}^{centr}_i &= - \sum_{k}^{K_N} \left(\boldsymbol{r}_i^k - \boldsymbol{r}_0^k \right) \otimes \boldsymbol{F}_i^k \label{eq:Centroid_Stress_Tensor} \\
    \boldsymbol{r}_0^k &= \sum_{j=1}^{N} p_j^k \boldsymbol{r}_j^k \label{eq:Centroid_Position}
\end{align}
In the above expression, $\otimes$ denotes the tensor product and $\boldsymbol{r}_0^k$ is the weighted geometric center (e.g. \textit{`centroid'}) of group $k$. Eq.~(\ref{eq:Centroid_Stress_Tensor}) is formally exact when all atoms vibrate strictly within the control volume $V$~\cite{Surblys2019}, which is in principle true for solid systems. Matsubara \textit{et al.} have in fact provided an exact formula for the volume-averaged HF that employed proper weighting coefficients so that it holds even when this conditions doesn't hold, and have made the modified LAMMPS source code publicly available, but only specifically for the Tersoff and AIREBO potentials~\cite{Matsubara2020}.

The average macroscopic value for $\boldsymbol{J}$ is found by time averaging, which should be zero for a system in equilibrium, while it must equal the flux calculated from the energy exchange rate of the thermostats in NEMD simulations (see Section~\ref{sec:NEMD}). The centroid formulation has been used in a previous work by one of the authors for many-body potentials where the potential energy was equally distributed among all atoms in each $k$-group~\cite{Surblys2019}. Eqs.~(\ref{eq:Centroid_Stress_Tensor}) and (\ref{eq:Centroid_Position}) are generalizations for arbitrary weights $p_i^k$. \\

\begin{figure*}[!tbp]
    \begin{center}
        \includegraphics[width=\textwidth]{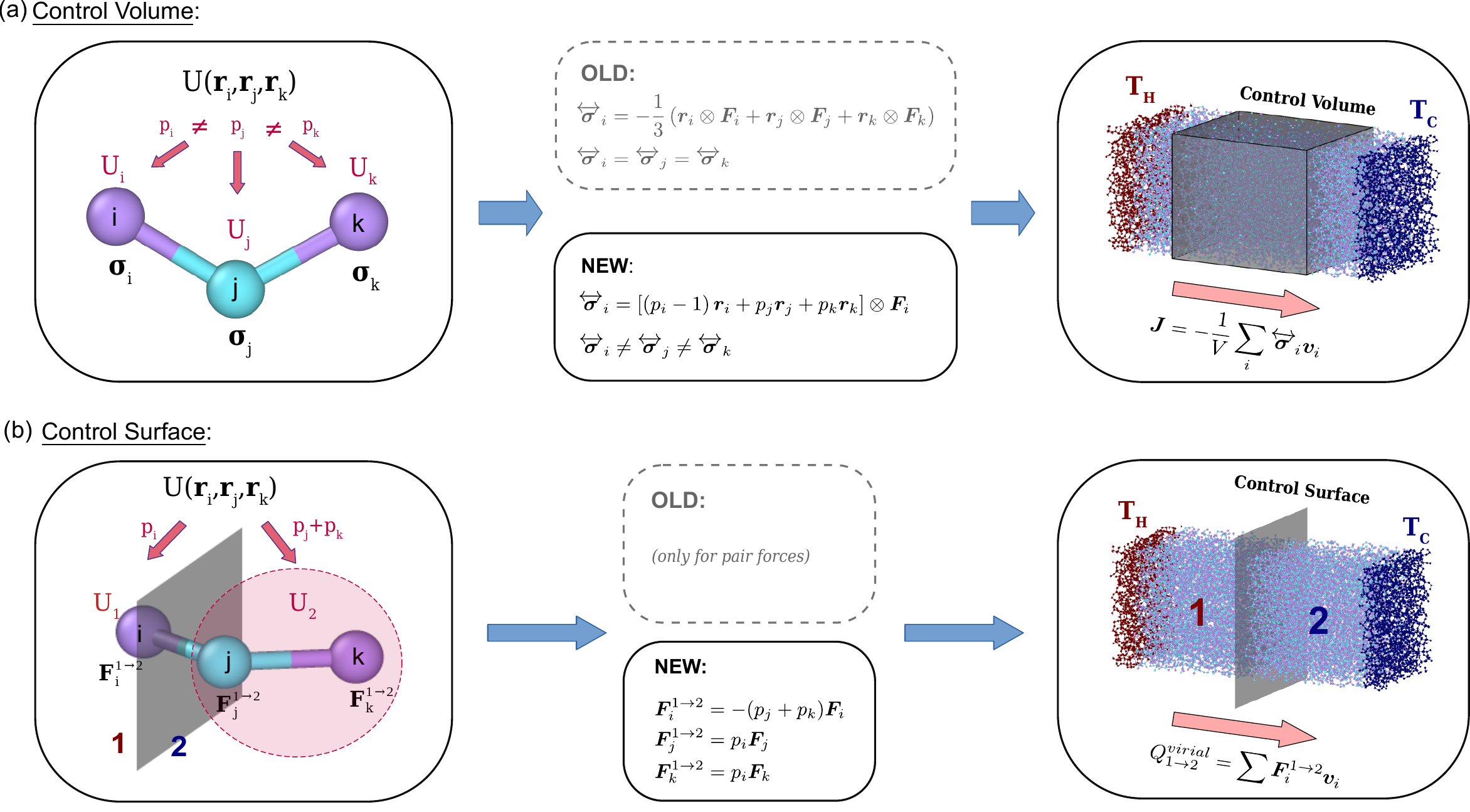}
        \caption{\label{fig:Visual_Abstract} Visualization of the HF calculation workflow for a 3-body interaction with arbitrary weights $p_i$. (a) Average HF in a Control Volume: the 3-body potential energy $U$ is distributed to $i$, $j$, $k$ according to their weights $p_i$ (left) and the atomic stress tensors are calculated (middle) according to the `old' group formula in eq.~(\ref{eq:LAMMPS_Stress_Tensor}) (gray) and the centroid formula from eqs.~(\ref{eq:Centroid_Stress_Tensor}) and~(\ref{eq:Centroid_Position}) (black). The HF is calculated in a control volume for both cases from eq.~(\ref{eq:HeatFlux_virial}) (right) (b) HF across a Control Surface: $U$ is distributed to regions $1$ and $2$ according to the weights $p_i$ of the atoms belonging to them (left) and the forces $\boldsymbol{F}_i^{1 \leftarrow 2}$ are calculated (middle) accordingly from eq.~(\ref{eq:F12}) (black). No previous such implementation exists except for pair forces (gray). $Q^{\text{virial}}_{1 \rightarrow 2}$ is then calculated from eq.~(\ref{eq:HeatCurrent_Torii}) (right), from which other quantities can be derived, such as $Q(\omega)$. Although bulk a-SiO$_2$ was chosen for visualization, the methodology is applicable to any arbitrary system.}
    \end{center}
\end{figure*}

\noindent\textbf{Atomic Stress Tensor (Group):} However, in the widely used LAMMPS code the volume-averaged heat flux is calculated using a different formula for the atomic stress tensor, given by~\cite{Plimpton1995}
\begin{equation}
\overleftrightarrow{\boldsymbol {{\sigma}}}_i^{\text{group}} = - \sum_{k=1}^{K} \frac{1}{N_k} \sum_{j \in k} \boldsymbol{r}_j^k \otimes \boldsymbol{F}_j^k \label{eq:LAMMPS_Stress_Tensor}
\end{equation}
where $N_k$ is the number of atoms participating in interaction group $k$. This formula, identically true to eq.~(\ref{eq:Centroid_Stress_Tensor}) for pair potentials ($N_k$=2), has been shown to yield incorrect results of thermal conductivity in a variety of many-body potentials~\cite{Surblys2019, Boone2019, fan2015force}, with the thermal conductivity of 1L graphene with Tersoff from eq.~(\ref{eq:LAMMPS_Stress_Tensor}) being almost 3 times lower than the one calculated from a formally exact formula by Fan~\textit{et al}~\cite{fan2015force}, which was shown to be equivalent to the centroid formula~(\ref{eq:Centroid_Stress_Tensor})$-$(\ref{eq:Centroid_Position})~\cite{Hamakawa2025}. The heat flux obtained from the atomic stress tensor given by eqs.~(\ref{eq:Centroid_Stress_Tensor}) and~(\ref{eq:Centroid_Position}) will henceforth be called \textit{`centroid HF'}, while the one obtained from eq.~(\ref{eq:LAMMPS_Stress_Tensor}) will be called \textit{`group HF'}, in order to maintain the terminology of the previous works~\cite{Surblys2019,Plimpton1995}. A schematic representation of the calculation of both the centroid and the group HF in a control volume for a 3-body interaction is shown in Fig.~\ref{fig:Visual_Abstract}(a).

\subsection{Heat Current through a Control Surface}
\noindent\textbf{Torii Formula:} A more formally exact way to determine the HF is by calculating the heat current flowing through a control surface of the system, which separates the computational cell into two regions, henceforth labeled $1$ and $2$. If we represent the heat current flowing from region $1$ to region $2$ by $Q_{1 \rightarrow 2}$, again using different notation Torii \textit{et al}. showed that the virial part of $Q_{1 \rightarrow 2}$ can be expressed as~\cite{Torii2008}
\begin{align}
    Q^{\text{virial}}_{1 \rightarrow 2} = \sum_{k}^{K_N} \sum_{i,j \in k \atop i>j}   \left(p_j^k \boldsymbol{F}_i^k \cdot \boldsymbol{v}_i^k - p_i^k \boldsymbol{F}_j^k \cdot \boldsymbol{v}_j^k \right)& \nonumber \\ 
     \times \left( H_i - H_j \right)& \label{eq:HeatCurrent_Torii}
\end{align}
The symbol $H_i$ is a Heaviside step function defined as
\begin{equation}
H_i = \begin{cases}
    1 & \text{if } i \in 2 \\
    0 & \text{otherwise}
\end{cases}
\end{equation}
and this practically means that the factor $\left( H_i - H_j \right)$ is equal to $(-1)$ when $i \in 1$ and $j \in 2$, equal to 1 when $i \in 2$ and $j \in 1$, and 0 when both atoms are in the same region. It can thus be thought of as a `directional ordering' of atoms $j$ and $i$ across the control surface, from region $1$ to region $2$ respectively.\\

\noindent\textbf{Per-atom contribution:} In an analogous manner to the expression of the HF from the atomic stress tensor, by noting the resemblance of eqs.~(\ref{eq:HeatCurrent_Torii}) and (\ref{eq:HeatFlux_Torii_V}), one can re-write $Q^{\text{virial}}$ as
\begin{equation}
    Q^{\text{virial}}_{1 \rightarrow 2} = \sum_i \boldsymbol{F}^{1 \rightarrow 2}_i \cdot \boldsymbol{v}_i \label{eq:HeatCurrent_Torii_Virial}
\end{equation}
where the term $\boldsymbol{F}^{1 \rightarrow 2}_i$ is defined as
\begin{align}
    \boldsymbol{F}^{1 \rightarrow 2}_i &=   \sum_{k}^{K_N} \left(H_i^k - H_0^k \right) \boldsymbol{F}_i^k \label{eq:F12} \\
    H_0^k &= \sum_{j \in k} p_j^k H_j \label{eq:H0k}
\end{align}

The quantity $Q^{1 \rightarrow 2}_i = \boldsymbol{F}^{1 \rightarrow 2}_i \cdot \boldsymbol{v}_i$ can then be considered as a per-atom contribution to the total heat current $Q^{\text{virial}}_{1 \rightarrow 2}$. If one defines $U^k_1$ and $U^k_2$ as the part of the potential energy $U^k$ of the interaction group $k$ attributed to region $1$ and $2$ respectively, it follows that $H_0^k=U^k_2/U^k$. This can be interpreted as the percentage of $U^k$ attributed to region $2$. Subsequently, the factor $\left( H_i^k - H_0^k \right)$ is equal to $U^k_1/U^k$ when $i \in 2$ and $-U^k_2/U^k$ when $i \in 1$. In other words, eqs.~(\ref{eq:HeatCurrent_Torii_Virial})-(\ref{eq:H0k}) show that for each atom participating in a many-body interaction, the percentage of the many-body forces responsible for transferring mechanical power from region $1$ to region $2$ is equal to the percentage of the interaction potential energy $U^k$ attributed to the atoms from the other region. In Fig.~\ref{fig:Visual_Abstract}(b), we provide a graphical representation of a 3-body potential energy being distributed to atoms $i$, $j$ and $k$ with weights $p_i$, $p_j$ and $p_k$ respectively. If atom $i$ belongs to region $1$ and atoms $j$ and $k$ to region $2$, then for atoms $j$ and $k$ the force $\boldsymbol{F}^{1 \rightarrow 2}$ is equal to $p_i\boldsymbol{F}_j$ and $p_i\boldsymbol{F}_k$ respectively, while for atom $i$ it is equal to $-(p_j+p_k)\boldsymbol{F}_i$.

This method for calculating the heat current (and by extension the HF) is particularly useful for various reasons. First of all, it allows the calculation of the exact heat flux passing through a material interface, in contrast to the volume-averaged HF from eq.~(\ref{eq:HeatFlux_Torii_V}), which only gives an average value and is in fact the integral of eq.~(\ref{eq:HeatCurrent_Torii}) along an axis perpendicular to the control surface~\cite{Torii2008}. It is not subject to any approximations and thus always strictly correct. Another major advantage is that $Q_{1 \rightarrow 2}$ is defined between groups of atoms and not a particular geometrical interface, and thus eq.~(\ref{eq:HeatCurrent_Torii_Virial}) is not restricted to a planar control surface, but can be used for the heat flow between any two groups of atoms. This can prove particularly useful in complex interface geometries, or to isolate specific HF components, such as the current passing through the membrane from the current leaking through the substrate in a supported 2D membrane, as the thermostats only provide the total HF.

\subsection{Spectral Decomposition of the Heat Flux}
We thus saw that the (average) virial heat current flowing across a control surface, as defined in eq.~(\ref{eq:HeatCurrent_Torii_Virial}), can be expressed as a sum of instantaneous atomic contributions, while at the same time, in order to ensure ergodicity, the true instantaneous heat current $Q_{1 \rightarrow 2}(t)$ measured in a system can be a statistical ensemble average over multiple simulation runs with different initial conditions, thus the current at time $t$ can be expressed as
\begin{equation}
   Q_{1 \rightarrow 2}(t) = \sum_i \langle\boldsymbol{F}^{1 \rightarrow 2}_i(t) \cdot \boldsymbol{v}_i(t) \rangle \label{eq:Q12_t}
\end{equation}
where the angular brackets $\langle \ldots \rangle$ denote an ensemble average. At steady state, the current fluctuates around its time-average value, which corresponds to the macroscopic value that becomes available to experimental measurements:
\begin{equation}
    \bar{Q}_{1 \rightarrow 2} = \lim_{\tau\rightarrow \infty} \ \frac{1}{\tau} \int_{-\tau/2}^{+\tau/2} Q_{1 \rightarrow 2}(t') \ dt' \label{eq:Q12_average}
\end{equation}
where $\tau$ here represents the observation time. One may also be interested to spectrally decompose the current $\bar{Q}_{1 \rightarrow 2}$ into contributions from modes with different frequencies $\omega$, for example in order to study which phonon spectral regions contribute most to the heat current. One can then define the Spectral Decomposition of the Heat Current (SDHC) $\bar{Q}_{1 \rightarrow 2}(\omega)$ in a way that it satisfies:
\begin{equation}
    \bar{Q}_{1 \rightarrow 2} = \int_{0}^{\infty} Q_{1 \rightarrow 2}(\omega) \ d\omega \label{eq:Q12_SD}
\end{equation}
One then may observe that the instantaneous current $Q_{1 \rightarrow 2}(t)$ is of the form $\sum_i \boldsymbol{A}_i(t) \cdot \boldsymbol{B}_i (t)$. By drawing an analogy with signal theory, where the average value of a composite, generally complex signal $C(t) = A^*(t) B(t)$ can be spectrally decomposed by exploiting Plancherel's theorem, which states that $\int_{-\infty}^{+\infty} A^*(t) B(t) \ dt \nonumber = \int_{-\infty}^{+\infty} \hat{A}^*(\omega)  \hat{B}(\omega) \ d\omega $, where $\hat{A}(\omega)$ denotes the Fourier transform of $A(t)$. In the case where $A(t)$ and $B(t)$ are both real functions, as is the case with $\boldsymbol{F}^{1 \rightarrow 2}_i$ and $\boldsymbol{v}_i$, it holds that $\hat{A}(-\omega) = \hat{A}^*(\omega)$, therefore it is possible to restrict the integration to positive frequencies only, allowing one to write $\int_{-\infty}^{+\infty} A^*(t) B(t) \ dt \nonumber = \int_{0}^{\infty} Re\{\hat{A}^*(\omega)  \hat{B}(\omega)\} \ d\omega$

By combining eqs.~(\ref{eq:Q12_t}), (\ref{eq:Q12_average}) with Plancherel's theorem, one can express the SDHC as
\begin{equation}
    Q_{1 \rightarrow 2}(\omega) = \frac{2}{\tau_{total}}\sum_i Re\{\langle \hat{\boldsymbol{F}}^{1 \rightarrow 2}_i(\omega) \cdot \hat{\boldsymbol{v}}^*_i(\omega) \rangle \} \label{eq:SDHC}
\end{equation}
where $\tau_{total}$ is the total simulation time for each simulation in the ensemble. This is in fact nothing more than the Cross-Spectral Density (CSD) of the velocities $\boldsymbol{v}_i$ and the forces $\boldsymbol{F}^{1 \rightarrow 2}_i$ acting on the atoms $i$ from accross the cross-section. Furthermore, by performing various projections, such as restricting the summation $\sum_i$ to specific atom species, considering only specific cartesian components for the vector product $\boldsymbol{F}^{1 \rightarrow 2}_i \cdot \boldsymbol{v}_i$, or restricting the first sumation in eq.~(\ref{eq:F12}) to specific interaction terms, one can study the contribution of specific atom species, phonon polarizations, or specific many-body terms to the heat current respectively, and by extent, to thermal conductivity. 

As an important remark, the SDHC calculated from eq.~(\ref{eq:SDHC}) is equivalent to the SDHC calculated through the use of the Fourier transform of the cross-correlation function $K_{ij}(\tau)= \frac{1}{2}\langle \boldsymbol{F}_{ij}(\tau) \cdot \boldsymbol{v}_i(0) - \boldsymbol{F}_{ji}(\tau) \cdot \boldsymbol{v}_j(0)\rangle$, as has been previously established by the works of S\"a\"askilahti et al.~\cite{Saaskilahti2014, Saaskilahti2015, Saaskilahti2016}. This is because at steady state the current $Q_{1 \rightarrow 2}(t)$ is stationary, and thus from the Wiener-Khinchin theorem it follows that the Fourier transform of $K(\tau)$ is equal to $Q_{1 \rightarrow 2}(\omega)$ in eq.~(\ref{eq:SDHC})~\cite{Blackman1959}. In this existing framework, the forces $\boldsymbol{F}_{ij}$ are also calculated either by the harmonic approximation for many-body potentials, or in an exact manner for pair potentials only~\cite{Saaskilahti2015,Saaskilahti2016}, as the definition of the many-body forces used in this framework is not convenient for general many-body potentials. Eq.~(\ref{eq:SDHC}) is also of complexity $\mathcal{O}(NlogN)$, while the calculation of $K(\tau)$ is of complexity $\mathcal{O}(N^2)$, where $N$ is the number of time samples. The use of eq.~(\ref{eq:SDHC}) can thus significantly speed up calculations.

\section{\label{sec:Results}Results}
In order to validate our methodology, we have performed comparative tests via both Equilibrium and Non-Equilibrium MD simulations. Some examples of the spectral decomposition of the NEMD heat current $Q(\omega)$ are also shown. The simulations investigated a variety of test systems, including both 2D and 3D systems, using the Tersoff and the Stillinger-Weber (SW) potentials, which both contain 2-body and 3-body terms. Concerning 2D systems, we investigated monolayer graphene (1LG), monolayer hBN (hBN) and some monolayer Transition Metal Dichalcogenides (TMD's) such as MoS$_2$, MoSe$_2$, MoTe$_2$, WS$_2$, WSe$_2$ and WTe$_2$. For the 3D systems we investigated crystalline Silicon (c-Si) and amorphous Silica (a-SiO$_2$). For all the above systems we used interatomic potential parameters already available in LAMMPS: the reparametrized Tersoff potential for 1LG~\cite{Lindsay2010} and hBN~\cite{Kinaci2012BN}, and SW with parameters by Jiang et al~\cite{Jiang2019} for the TMD's. For c-Si we used for comparison both the Tersoff~\cite{Tersoff1988} and the SW~\cite{Stillinger1985} potentials, while for a-SiO$_2$ the Tersoff~\cite{Munetoh2007} potential was employed. Numerical results are provided in Table~\ref{tab:numerical_results}. The computational details for all simulations performed in this work are presented in the Supplementary Material: an overview of the systems used is shown in Fig.~S1, while example calculations with GK and NEMD simulation data are presented in Figs.~S2 and S3, respectively.
\begin{figure}[!t]
    \includegraphics[width=0.95\columnwidth]{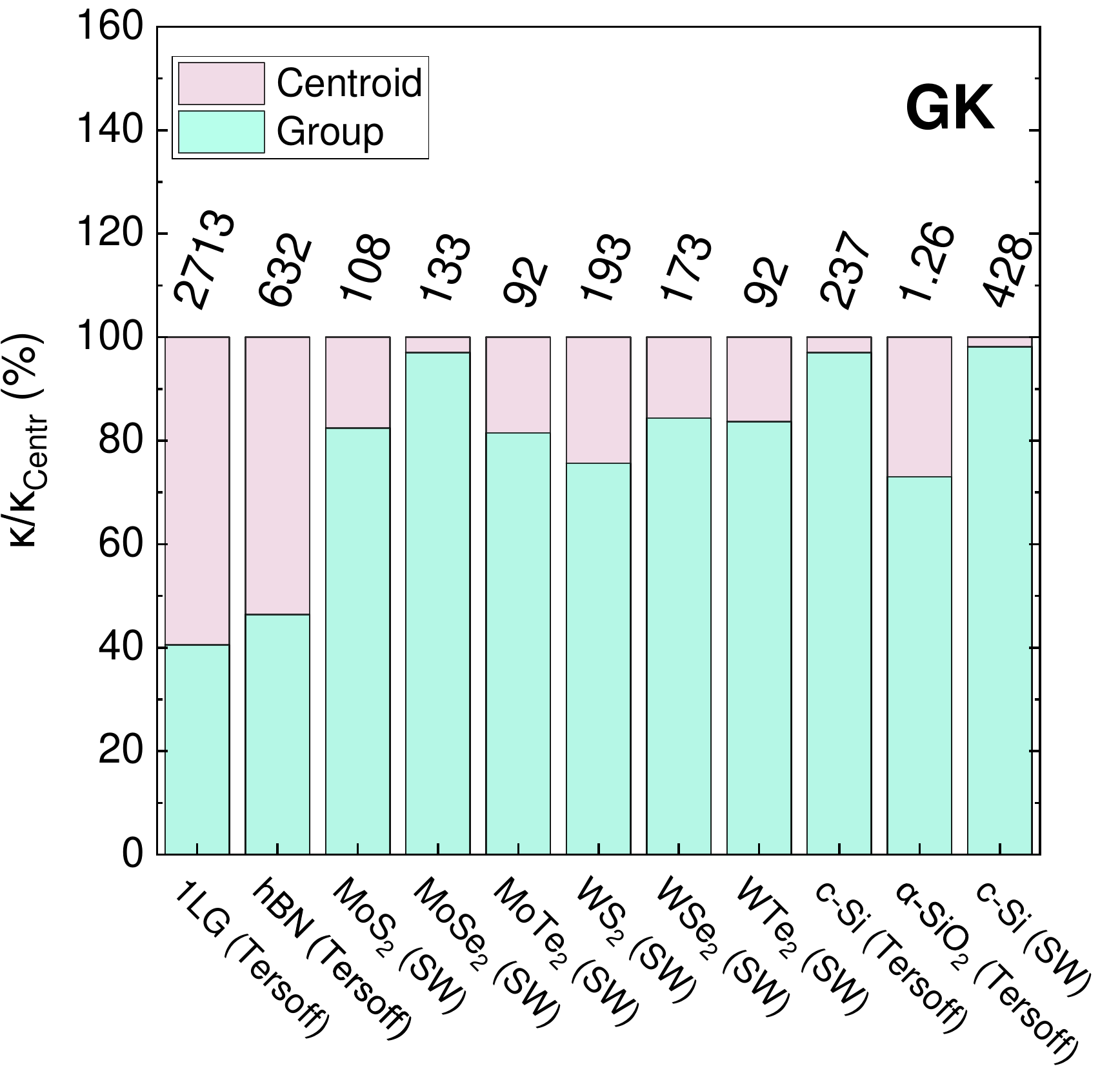}
    \caption{\label{fig:GK_comp} Comparative results with the Green-Kubo method. The calculated thermal conductivity $\kappa$ of various 2D and 3D systems using GK and $\boldsymbol{J}$ calculated from the centroid formula of eq.~(\ref{eq:Centroid_Stress_Tensor}) (\textit{in pink}) and the group formula of eq.~(\ref{eq:LAMMPS_Stress_Tensor}) (\textit{in blue}). $\kappa_{group}$ is shown as a percentage of $\kappa_{centr}$. For the centroid formula, the absolute values of $\kappa$ (in $W/mK$) are also displayed. In all cases, the uncertainties were of the order of 2-5\%}
\end{figure}
\subsection{Green-Kubo simulations} 

We start our validation by calculating the thermal conductivity $\kappa$ of the test systems using the Green-Kubo method, using both the centroid and the group formulation of the HF. All systems were kept in equilibrium at room temperature $T$=300K under periodic boundary conditions to simulate an infinite bulk system. The results are presented in Fig.~\ref{fig:GK_comp}, where we compare the values of $\kappa$ obtained with the centroid and the group formulation, henceforth labeled as $\kappa_{centr}$ and $\kappa_{group}$ respectively. The results show that for the systems and potentials investigated, the group atomic stress tensor systematically underestimates $\kappa$ compared to the centroid formulation, with the most pronounced difference being for 1LG, where $\kappa$ of monolayer graphene is underestimated by 60\% by the existing implementation of the heat flux in LAMMPS. More precisely, $\kappa_{group}$ (1100 W/mK) is $\sim$40\% of $\kappa_{centr}$ (2713 W/mK). The centroid value of thermal conductivity for 1LG is in excellent agreement with the one calculated by Fan \textit{et al}.~\cite{fan2015force} using the Tersoff potential and a formula equivalent to the centroid HF. For hBN $\kappa_{group}$ is significantly lower ($\approx$45\%) compared to $\kappa_{centr}$. For the TMD's the underestimation of $\kappa$ by the group stress tensor is in the range of 20\%. These findings are also in line with other previous works which have demontrated the systematic underestimation of $\kappa$ by the current HF formulation of LAMMPS for many-body potentials~\cite{Surblys2019, Boone2019}. Nonetheless it should be noted that this is not a priori the case; for example for c-Si, the two formulations give very similar results both under the Tersoff and SW potentials.

\begin{figure}[!t]
    \includegraphics[width=\columnwidth]{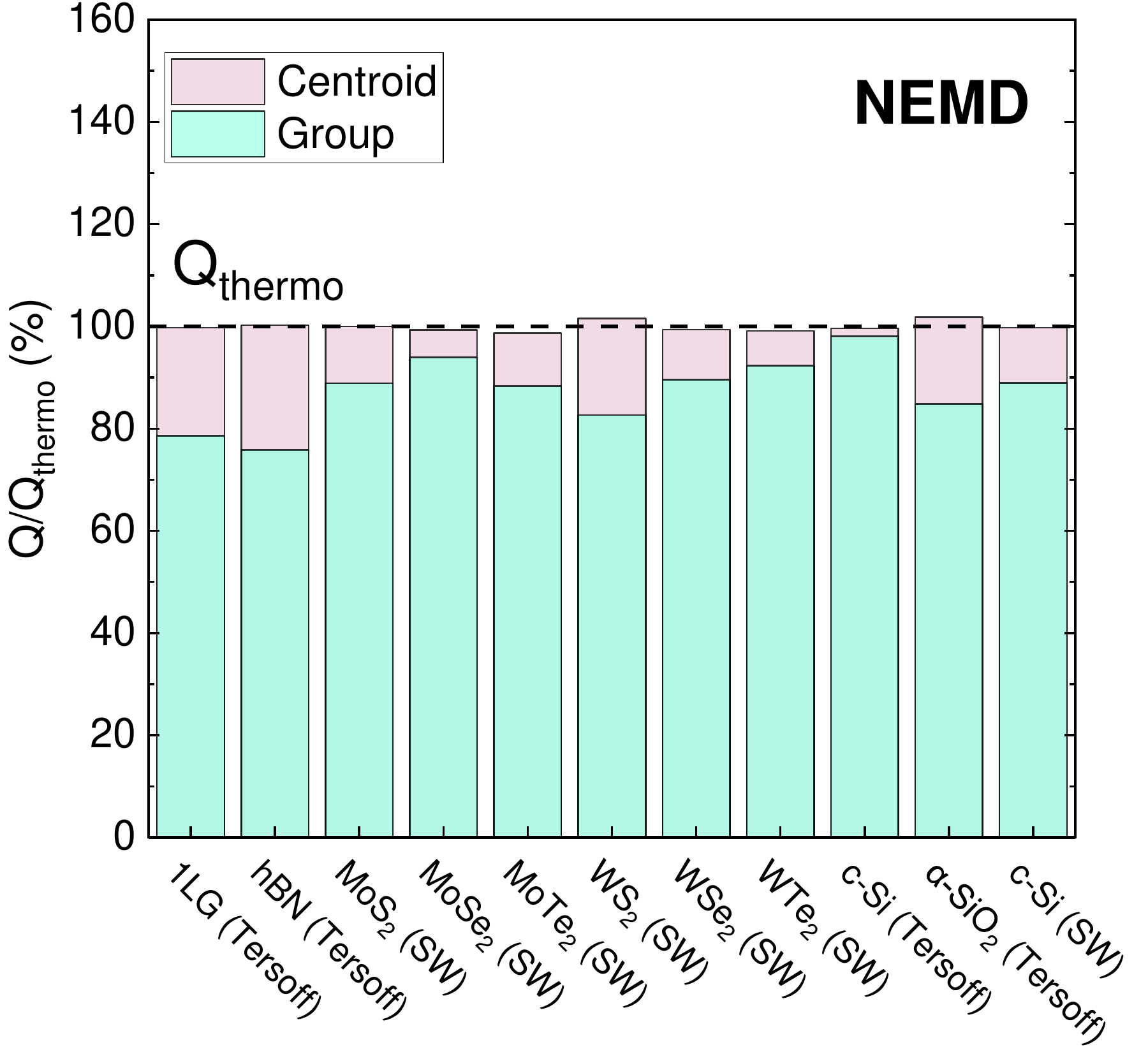}
    \caption{\label{fig:NEMD_comp} Comparative results with the NEMD method. The heat currents from the mean heat flux values calculated with the centroid formula of eq.~(\ref{eq:Centroid_Stress_Tensor}) ($Q_{centr}$, \textit{in pink}) and the group formula of eq.~(\ref{eq:LAMMPS_Stress_Tensor}) ($Q_{group}$, \textit{in blue}), shown as a percentage of the current $Q_{thermo}$ calculated from the energy exchange rate of the thermostats. For the values of $Q$ from the HF, the uncertainties were of the order of 2-5\%, while for $Q_{thermo}$ it was below 1\%.}
\end{figure}

\subsection{\label{sec:NEMD}NEMD simulations} 
We further validate our methodology by performing Non-Equilibrium Molecular Dynamics (NEMD) simulations to directly compute the average heat current from eq.~(\ref{eq:HeatFlux_virial}) under the group ($Q_{group}$) and the centroid atomic stress tensors ($Q_{centr}$). We compare these values with the current calculated from the energy exchange rate of the thermostats ($Q_{thermo}$), which is generally accepted as the most exact estimation of the heat current in NEMD simulations. In principle the current calculated from the atomic stress tensor and should be equal to $Q_{thermo}$ under steady state conditions, where the heat flux is stationary and constant along the thermal gradient direction. The results are presented in Fig.~\ref{fig:NEMD_comp}, where we compare $Q_{centr}$ and $Q_{group}$ with the heat bath current $Q_{therm}$. As a first remark, our results show the remarkable agreement between $Q_{centr}$ and $Q_{thermo}$, which serves to validate our centroid method implementation. The existing LAMMPS group formula exhibits again a similar tendency as in the Green-Kubo simulations, namely that it systematically underestimates the heat current, compared now to both the centroid formula and the thermostat current. The underestimation is again more pronounced for 1LG and hBN, but it is important to note that it is less pronounced than for the GK results. This can be attributed to the fact that for NEMD one calculates the time-average value of $J$, while for GK it is the integral of the autocorrelation function $\langle J(\tau) J(0) \rangle$, which is much more sensitive to differences in the instantaneous heat flux.

\begin{table}[!t]
  \caption{\label{tab:numerical_results} Numerical results of thermal conductivities for all systems mentioned in this work, calculated both with GK and NEMD, along with the system dimensions. For the 2D systems $l_z$ was chosen as the bulk interlayer distance. For the NEMD simulations, $\Delta T$=120~K and for all simulations the average temperature was $T_0$=300~K. For the GK results, the uncertainty is $\sim$5\%, while for NEMD it is less than 1\%.}
  \begin{ruledtabular}
      \begin{tabular}{llllll}
      \multirow{2}{*}{\textbf{System}} &\multirow{2}{*}{\textbf{Method}} &\textbf{$l_x$}  & \textbf{$l_y$} & \textbf{$l_z$}  & $\kappa$ \ \\
      & &\multicolumn{3}{c}{(nm)} \ & (W/mK)\\ \hline
       1LG (Tersoff)& GK& 10& 10 & 0.335 & 2790  \\
       1LG (Tersoff)& NEMD& 100& 10 & 0.335 & 564 \\
       hBN (Tersoff)& GK& 10& 10 & 0.335 & 632  \\
       hBN (Tersoff)& NEMD& 100& 10 & 0.335 & 262 \\   
       MoS$_2$ (SW)& GK& 10& 10 & 0.615 & 108  \\
       MoS$_2$ (SW)& NEMD& 100& 10 & 0.615 & 20.5 \\   
       MoSe$_2$ (SW)& GK& 10& 10 & 0.644 & 127  \\
       MoSe$_2$ (SW)& NEMD& 100& 10 & 0.644 & 22.6 \\
       MoTe$_2$ (SW)& GK& 10& 10 & 0.690 & 82  \\
       MoTe$_2$ (SW)& NEMD& 100& 10 &  0.690 & 11.5 \\
       WS$_2$ (SW)& GK& 10& 10 & 0.760 & 156.1  \\
       WS$_2$ (SW)& NEMD& 100& 10 & 0.760 & 16.1  \\ 
       WSe$_2$ (SW)& GK& 10& 10 & 0.645 & 165  \\
       WSe$_2$ (SW)& NEMD& 100& 10 & 0.645 & 25 \\ 
       WTe$_2$ (SW)& GK& 10& 10 & 0.615 & 90  \\
       WTe$_2$ (SW)& NEMD& 100& 10 & 0.615 & 10.5 \\  
       \hline
       c-Si (Tersoff)& GK& 5& 5 & 5 & 237  \\
       c-Si (Tersoff)& NEMD& 30& 5 & 5 & 17 \\  
       a-SiO$_2$ (Tersoff)& GK& 5& 5 & 5& 1.26  \\
       a-SiO$_2$ (Tersoff)& NEMD& 30& 5 & 5 & 1.22 \\
       c-Si (SW)& GK& 5& 5 & 5& 430  \\
       c-Si (SW)& NEMD& 30& 5 & 5 & 13.1 \\  
       \end{tabular}
   \end{ruledtabular}
\end{table}

\subsection{Spectral Decomposition of the Heat Current}
\begin{figure}[!t]
    \includegraphics[width=0.8\columnwidth]{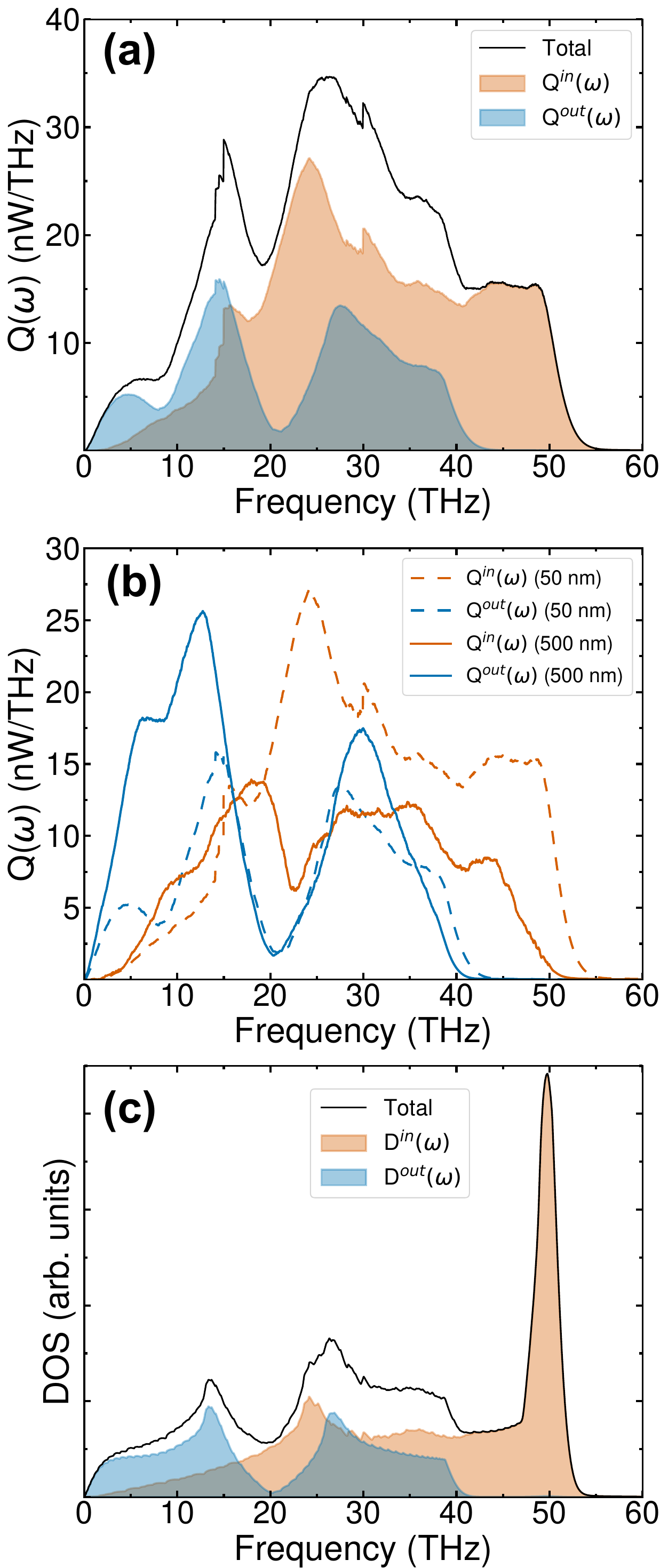}
\caption{\label{fig:1L_SDHC} The spectral decomposition $Q(\omega)$ for 1LG for different system lengths. (a) $Q(\omega)$ for 1LG of length $L$=50~nm (\textit{in black}). The in-plane component $Q^{\text{in}}$ (\textit{in red}) and the out-of-plane component $Q^{\text{out}}$ (\textit{in blue}) are also shown. (b) $Q^{\text{in}}$ (\textit{in red}) and $Q^{\text{out}}$ (\textit{in blue}) components are shown for different system lengths: $L$=50~nm (\textit{solid lines}) and $L$=500~nm (\textit{dashed lines}). (c) The phonon DOS of 1LG (\textit{in black}), along with the in-plane (\textit{in red}) and out-of-plane (\textit{in blue}) contribution, normalized to unit total area.}
\end{figure}

\begin{figure}[!t]
    \includegraphics[width=0.8\columnwidth]{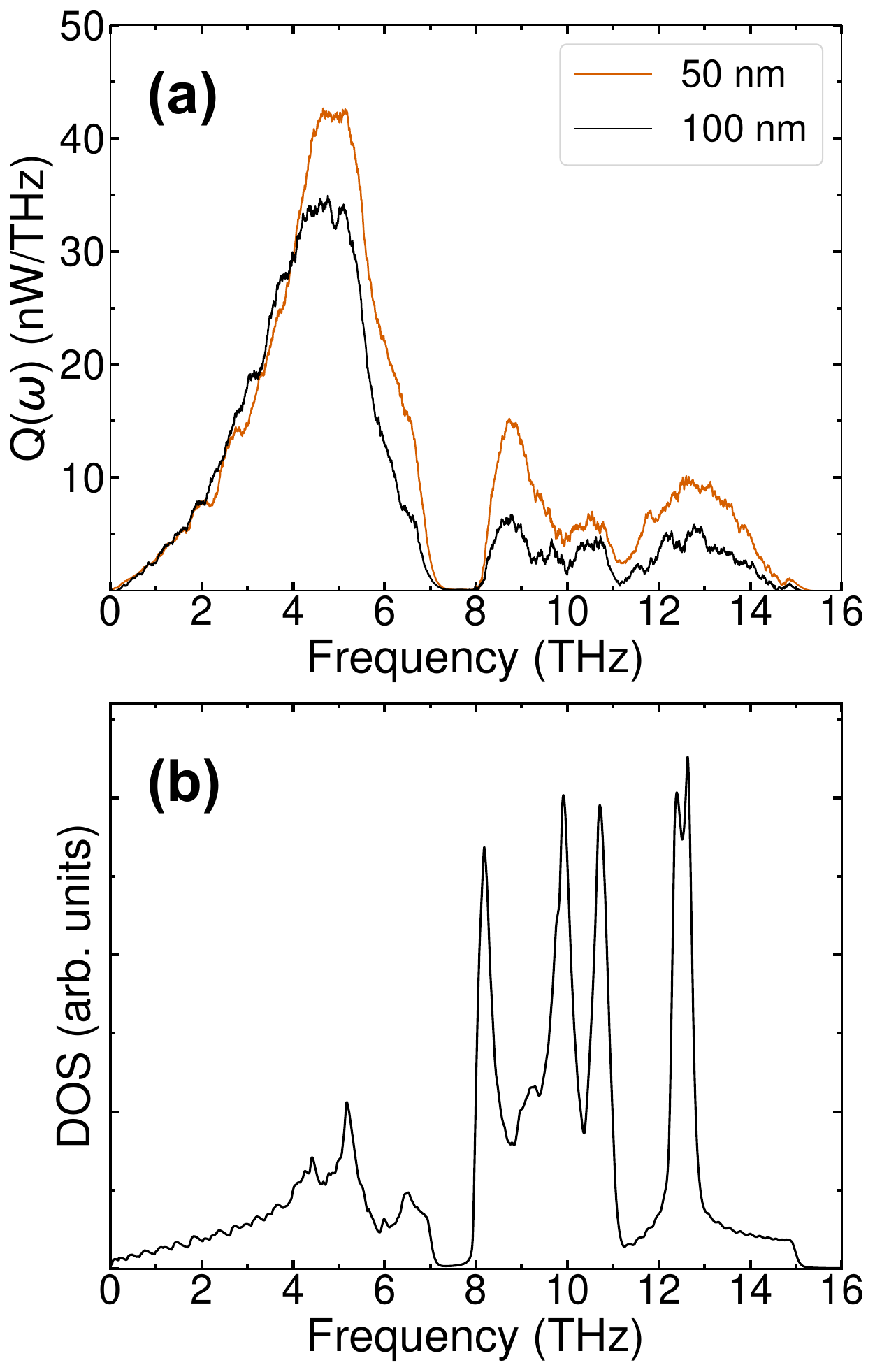}
    \caption{\label{fig:MoS2_SDHC} The spectral decomposition of the heat current $Q(\omega)$ for MoS$_2$ with different system lengths (a) $Q(\omega)$ for $L$=50~nm (\textit{red}) and $L$=500~nm (\textit{black}). (c) The total phonon DOS of MoS$_2$, normalized to unit total area.}
\end{figure}

As a final validation of our methodology, we have also performed NEMD simulations for 1LG and hBN to calculate the instantaneous virial heat current $Q_{1 \rightarrow 2}^{virial}(t)$ across a control plane in the middle of the system (as in Fig.~\ref{fig:Visual_Abstract}(b), left), as well as its spectral decomposition $Q(\omega)$. The formulas presented in eqs.~(\ref{eq:HeatCurrent_Torii_Virial}) and~(\ref{eq:SDHC}) were used. Both systems are solids, therefore the convective part of the heat current can be neglected. The phonon Density of States (DOS) of both systems was also calculated from the Fourier transform of the velocities $\boldsymbol{v}_i$ of the atoms around the cross-section
\begin{equation}
D(\omega)=\frac{1}{3k_B T}\sum_i m_i \langle \hat{\boldsymbol{v}}_i(\omega) \cdot \hat{\boldsymbol{v}}^*_i(\omega) \rangle \label{eq:DOS}
\end{equation}
as described in detail in refs~\cite{Poulos2024, Koukaras2015, Chakraborty2018}. Here $m_i$ denotes the mass of atom $i$, $T$ is the temperature and $k_B$ the Boltzmann constant. In all cases though, $D(\omega)$ is normalized to unit total area. Two different system lengths were investigated for each system, in order to show the dependence of $Q(\omega)$ on the system size. The results for 1LG and MoS$_2$ are presented in Figs.~\ref{fig:1L_SDHC} and~\ref{fig:MoS2_SDHC} respectively. In all the above cases, the time-averaged currents $Q_{1 \rightarrow 2}^{virial}$, $Q_{centr}$ and $Q_{thermo}$ were found to be in excellent agreement at steady state.

\noindent\textbf{Graphene:} Starting with 1LG, one can first note that due to graphene's pure 2D nature, in-plane phonon modes (LA, TA, LO, TO) have purely $x$- and $y$-components in their eigenvectors, while out-of-plane modes (ZA, ZO) have purely $z$-components, assuming a graphene sheet parallel to the $xy$-plane. This way, one is able to perform a polarization decomposition of $Q(\omega)$ into in-plane and out-of-plane components by defining:
\begin{align}
    Q^{\text{in}}(\omega) &= \sum_i F^{1 \rightarrow 2}_{i,x} v_{i,x} + F^{1 \rightarrow 2}_{i,y} v_{i,y}\\
     Q^{\text{out}}(\omega) &= \sum_i F^{1 \rightarrow 2}_{i,z} v_{i,z}
\end{align}
A similar decomposition can be done for the DOS by keeping the $x$- and $y$-components and the $z$-component for $D^{\text{in}}(\omega)$ and $D^{\text{out}}(\omega)$ respectively in eq.~(\ref{eq:DOS}). In Fig.~\ref{fig:1L_SDHC}(a) we present $Q^{\text{in}}(\omega)$, $Q^{\text{out}}(\omega)$ and the total $Q(\omega)$ for 1LG of length L=50~nm, which shows a renarkable agreement with a similar earlier work by Fan et al. (fig.~4c)~\cite{Fan2017SDHC}, although a direct comparison is not possible, as the system size is not mentioned in that work. From Fig.~\ref{fig:1L_SDHC}(b) we can observe that for 1LG the dependence of $Q(\omega)$ on the system size is indeed quite strong. More interestingly, the two polarization components are affected in opposite ways: the in-plane contribution reduces significantly with $L$, especially above 20~THz which is a spectral region corresponding to the LO/TO optical and low-$\lambda$ LA/TA acoustical modes. On the other hand the out-of-plane contribution increases significantly with $L$, especially below 20~THz, which corresponds to the ZA acoustical modes. In fact, for $L$=500~nm the contribution of the out-of-plane modes to the total current becomes dominant (53\% vs 31\% at 50~nm), which is determined by integrating each polarization component.

In a previous work we have also shown, using the k-space Velocity Autocorrelation Sequence (kVACS) method under MD and the same interatomic potential, that in pristine 1LG the ZA and ZO modes have noticeably higher lifetimes $\tau$ than the in-plane modes~\cite{Poulos2024}. This increased $\tau$ for the ZA may imply higher intrinsic phonon Mean Free Paths (MFP), which can explain their dominant contribution in the thermal conductivity at higher $L$. For small systems their MFP is restricted from the system dimensions and thus thermal transport becomes ballistic. Assuming a definition of thermal conductivity as $\kappa=(Q/A)\cdot(L/\Delta T)$, where $A$ is the cross-sectional area and $\Delta T$ is the temperature difference, we can estimate $\kappa$ of 1LG for the two system lengths as $\kappa_{50}$= 120 W/mK and $\kappa_{500}$= 1070 W/mK. From the polarization decomposition we thus have $\kappa_{50}^{in}$= 83 W/mK and $\kappa_{50}^{out}$= 37 W/mK, while $\kappa_{500}^{in}$= 500 W/mK and $\kappa_{500}^{out}$= 570 W/mK. In other words, with a 10-fold increase in length, $\kappa$ increases by a factor of 9; $\kappa^{in}$ increases by 6, while $\kappa^{out}$ increases by a remarkable factor of 15. This almost linear increase of $\kappa^{total}$ with size indicates that the system is still in the ballistic regime even at 500~nm; previous works have shown that $\kappa$ of 1LG reaches a plateau at $\sim$15~$\mu$m with MD~\cite{Barbarino2015} or at $\sim$100~$\mu$m with BTE (and only 3$ph$ scattering contributions)~\cite{Fugallo2014}. The fast increase of $\kappa^{out}$ with $L$ then implies that this slow convergence can be attributed to the out-of-plane modes.

The largest dependence on system size is observed for the spectral region of the ZA modes: 34\% at 500~nm against 15\% at 50~nm. A previous work has also estimated the contribution of the ZA modes to the thermal conductivity of 1LG at $\sim$34\% at 9~$\mu$m, by exact solutions to the Boltzmann Transport Equation (BTE) under 3$ph$+4$ph$ scattering and based on the same Tersoff potential~\cite{Feng2018}. A more recent work using full ab-initio calculations has shown a much higher contribution of the ZA modes to 73\% to $\kappa$ of pristine 1LG at 300~K~\cite{Han2023}. The authors in this work implied that the Tersoff potential overestimates the 4$ph$ scattering of ZA modes, which has been shown to be the dominant scattering mechanism for ZA modes in 1LG due to graphene's plane reflection symmetry~\cite{Han2023}.

\noindent\textbf{MoS$_{\boldsymbol{2}}$:} As a final example, we consider the case of the spectral decomposition $Q(\omega)$ for monolayer MoS$_2$, for the first time to our knowledge, for two different system lengths, $L$=50~nm and $L$=500~nm, as shown in Fig.~\ref{fig:MoS2_SDHC}(a). In Fig.~\ref{fig:MoS2_SDHC}(b) we also show the total phonon DOS of MoS$_2$, where a clear gap is observed at $\sim$7~THz, which separates the acoustic modes below the gap from the optical modes above it. It is evident from the figure that $Q(\omega)$ is dominated by the acoustic modes even for lengths as small as 50~nm, which is in contrast to the case of 1LG. The acoustic phonon contribution to $Q$ is calculated by integrating $Q(\omega)$ below the gap, which gives $\sim$74\% at 50~nm and $\sim$84\% at 500~nm. The acoustical modes are thus already vastly dominant at 50~nm. By using the same definition for $\kappa$ as before, we can $\kappa_{50}$= 11.5 W/mK and $\kappa_{500}$= 17.5 W/mK for MoS$_2$. The acoustical and optical contributions to $\kappa$ then become $\kappa_{50}^{ac}$= 8.5 W/mK and $\kappa_{50}^{op}$= 3 W/mK respectively, while $\kappa_{500}^{ac}$= 14.7 W/mK and $\kappa_{500}^{op}$= 2.8 W/mK. In other words, at these length scales a doubling of the system size leads to a 50\% increase of $\kappa$ for MoS$_2$, which is significantly lower than the 9-fold increase for 1LG. At the same time, the acoustic phonon contribution $\kappa^{ac}$ increases by 73\% while the optical phonon contribution $\kappa^{op}$ decreases by 7\%. This can be attributed to the fact that the acoustic modes are already dominant at small $L$, so their MFP is not restricted by the system size, while the optical modes have much lower MFP due to their higher frequency and thus higher scattering rates.

It shall be noted that contrary to graphene, MoS$_2$ is not a purely 2D material, as it consists of a monolayer of Mo atoms sandwiched between two layers of S atoms; the in-plane (longitudinal and transverse) and out-of-plane (flexural) contributions cannot be distinguished simply by their cartesian components, as both have contributions in all three cartesian directions, especially for small-wavelength modes where a breathing motion of the unit cell occurs. Therefore, the distinction between in-plane and out-of-plane modes is not as straightforward as in the case of 1LG and for reasons of simplicity we have not performed a polarization decomposition of the heat current for this example.

\section{\label{sec:Conclusions}Conclusions}

In this work we have derived an exact framework for the calculation of the microscopic heat flux for MD in a control volume and across a control surface, as well its Spectral Decomposition, for arbitrary many-body potentials based on the theoretical work by Torii \textit{et al}.~\cite{Torii2008}. 
We have also implemented and tested this framework within the widely used open-source LAMMPS code, enabling the exact calculation of the heat flux and its spectral decomposition with MD. The tests included both Green-Kubo and Non-Equilibrium MD simulations on a variety of 2D and 3D test systems, including monolayer graphene (1LG), hBN, various TMD's, crystalline Silicon (c-Si) and amorphous Silica (a-SiO$_2$), using the widely used Tersoff and Stillinger-Weber potentials. 

Our results show that the existing HF implementation in LAMMPS systematically underestimates the thermal conductivity of these materials, especially for 1LG and hBN, and more strongly with Green-Kubo simulations than with NEMD. Our modifications lead to results for the thermal conductivity of 1LG with Tersoff that is in excellent agreement with a previous implementation of an exact HF formula for the Tersoff potential by Fan~\textit{et al.}~\cite{fan2015force}. In all NEMD simulations, the heat current calculated both from the control volume and the control surface methods is in perfect agreement with the thermostat current, which proves the consistency of our results. 

We have finally presented examples of the spectral decomposition of the heat current for 1LG and MoS$_2$, which shows a strong dependence on system size. For 1LG, the contribution of the out-of-plane modes to the total thermal conductivity was shown to become dominant at larger system sizes, with the ZA acoustic modes accounting for 35\% of the total thermal conductivity at 500~nm, highlighting the importance of these modes in thermal transport of graphene. For MoS$_2$, the acoustic phonon modes were found to contribute over 74\% to the heat current even at lengths as low as 50~nm, increasing with size. 

The modified LAMMPS source files for the Tersoff and SW potentials are available upon request and shall be made publically available in due time. The methodology presented in this work can be also readily implemented for other many-body potentials as well, which will be the subject of future work.

\section*{\label{sec:Acknowledgments} Acknowledgments}

A part of the calculations presented in this work was performed in the \textit{Jean-Zay} CPU array of the IDRIS High Performance Computing facilities (HPC) under the allocation 2024-AD010913913 made by GENCI. D.S acknowledges support by JSPS KAKENHI Grant Number JP24K07334. M.P would like to thank Ali Rajabpour, Bohayra Mortazavi and Julien El Hajj for fruitful discussions on existing implementations of heat flux calculations in LAMMPS.

\newpage

%

\end{document}